\documentclass[a4paper, 11pt]{article}
\usepackage{graphicx}
\usepackage{pdfpages}
\usepackage[utf8]{inputenc}
\usepackage{amsmath}
\usepackage{amsthm}
\usepackage{newtxtext,newtxmath}
\usepackage{natbib} 
\usepackage{bm}
\usepackage[unicode, psdextra, linkcolor=blue, citecolor=blue, colorlinks=true]{hyperref}
\usepackage{algorithm,algorithmic}
\usepackage{mathtools}
\allowdisplaybreaks
\usepackage{makecell, siunitx, multirow, threeparttable, caption, stackengine}
\usepackage{pdflscape}
\usepackage[nottoc]{tocbibind}
\usepackage[titletoc]{appendix}
\usepackage{afterpage,natbib,lipsum}
\usepackage[unicode, psdextra, linkcolor=blue, citecolor=blue, colorlinks=true]{hyperref}
\usepackage[top = 3.5cm, bottom = 3.5cm, left = 2.5cm, right = 2.5cm]{geometry}
\usepackage{accents}
\usepackage{xcolor}
\usepackage{authblk}
\usepackage{array}
\usepackage{bookmark, xpatch}
\usepackage[symbols,nogroupskip,sort=none,numberedsection=autolabel]{glossaries-extra}

\DeclareMathOperator*{\argmax}{arg\,max}

\newtheorem{assumption}{Assumption}
\newtheorem{theorem}{Theorem}

\title{Debiasing Sample Loadings and Scores in Exponential Family PCA for Sparse Count Data}
\author{Ruochen Huang}
\author{Yoonkyung Lee}
\affil{The Ohio State University}
\date{\vspace{-5ex}}

\begin{document}
\maketitle

\begin{abstract}
Multivariate count data with many zeros frequently occur in a variety of application areas such as text mining with a document-term matrix and cluster analysis with microbiome abundance data. Exponential family PCA \citep{collins2001generalization} is a widely used dimension reduction tool to understand and capture the underlying low-rank structure of count data. It produces principal component scores by fitting Poisson regression models with estimated loadings as covariates. This tends to result in extreme scores for sparse count data significantly deviating from true scores. We consider two major sources of bias in this estimation procedure and propose ways to reduce their effects. First, the discrepancy between true loadings and their estimates under a limited sample size largely degrades the quality of score estimates. By treating estimated loadings as covariates with bias and measurement errors, we debias score estimates, using the iterative bootstrap method for loadings and considering classical measurement error models. Second, the existence of MLE bias is often ignored in score estimation, but this bias could be removed through well-known MLE bias reduction methods. We demonstrate the effectiveness of the proposed bias correction procedure through experiments on both simulated data and real data. 
\vspace{8pt}
	

\end{abstract}

\section{Introduction\label{intro}}
Multivariate count data have increasingly emerged in numerous modern application domains such as text mining \citep{aggarwal2012introduction}, genetics \citep{pierson2015zifa,xu2021zero}, and recommendation systems \citep{koren2009matrix}. 
A notable characteristic of count data from those applications is sparsity, meaning that they contain many zeros. 
As a well-known example in music recommender systems, the million song dataset \citep{bertin2011million} provides information about users' listening history, covering a wide range of songs. Typically, a user might have a strong interest in one or a few categories of songs and rarely listens to all songs. Therefore, the proportion of zeros in the resulting user-song frequency matrix can easily reach more than $80\%$. In addition to sparsity, high dimensionality is another salient feature of count data in the aforementioned applications. To handle such data efficiently and model the associations among a large number of variables, we often reduce the data dimensionality by considering a low-dimensional latent structure.

Principal component analysis (PCA) \citep{jolliffe2002principal} has been a very effective dimension reduction tool. However, the implicit connection to a normal distribution makes this method improper for counts, which are typically modeled as a Poisson distribution. A series of efforts have been made to generalize PCA to non-Gaussian distributions, mostly exponential family distributions, including a Poisson distribution. Alternatively, non-negative matrix factorization \citep{lee1999learning} has been developed as an extension of PCA for non-negative data under square error loss with constraints that loadings and scores are non-negative.  \cite{collins2001generalization} proposed exponential family PCA as an extension of standard PCA for exponential family data under the generalized linear model framework. It accommodates a low-dimensional latent structure through the key assumption that the natural parameter matrix can be factorized into two low-rank matrices, namely, the principal component scores matrix and the loadings matrix. Inspired by Pearson's version of standard PCA, \cite{landgraf2020generalized} reformulated exponential family PCA by defining a specific low-rank form of the natural parameter matrix through projection of the saturated model parameters. There are also several approaches that modify the exponential family PCA based on the Bayesian framework by treating loadings or scores as random variables \citep{mohamed2008bayesian,li2010simple,acharya2015nonparametric, lin2021exponential}. Alternatively, some methods impose a low-rank structure on the mean parameters of exponential family distributions rather than natural parameters \citep{liu2018pca,chiquet2018variational,kenney2021poisson}. See \cite{smallman2022literature} for a more comprehensive review. 

Among the aforementioned approaches, exponential family PCA is a widely used method due to its straightforward formulation and easy-to-use algorithms for computation \citep{genpca}. Exponential family PCA treats loadings and scores as fixed parameters and estimates them jointly through the maximum likelihood method.
The joint estimation is computationally efficient but produces inconsistent estimators for loadings and scores under the classical asymptotic setting where the sample size $n$ goes to infinity while the number of features $d$ is fixed and not increasing with sample size.
This phenomenon is known as an ``incidental parameter problem" \citep{lancaster2000incidental}, where consistent estimators of structural parameters cannot be obtained under the existence of incidental parameters when we jointly estimate both types of parameters by maximizing the likelihood. In exponential family PCA, the loadings are the structural parameters whose size does not change with $n$ while the scores are the incidental parameters whose size grows with $n$. 
\cite{huang2023dissertation} theoretically showed that the population version of exponential family PCA for binary data is not Fisher consistent. Exponential family PCA only yields consistent estimators when both the sample size and feature dimension go to infinity \citep{wang2021deviance}, which corresponds to a large $n$ and large $d$ scenario. Concerns about the quality of estimated loadings and scores naturally arise when we apply exponential family PCA to count data where either $n$ or $d$ is not large enough.

This motivates us to explore the quality of estimated loadings and scores numerically through simulations
to find any evidence for modification of the estimators. Here we provide a summary of our observations from the simulation study (see Section~\ref{ex} for a detailed analysis of motivating simulation examples). 
We find that estimates of loadings often show some systematic patterns of bias. 
These nontrivial estimation errors cannot be explained by the sampling variability alone. Instead, there exist some clear patterns in the discrepancy between estimated loadings and true loadings such as inflation or deflation over a certain range of true loading values, which inspires us to presume that
\begin{equation}
    \mathrm{estimated~loading ~=~true~loading~+~bias~+~sampling~error}.
\end{equation}
In addition, estimated scores are also biased, due to the bias from loadings as well as the MLE bias involved in the joint maximum likelihood estimation procedure. 

We propose a bias correction procedure to improve the quality of estimated loadings and scores in exponential family PCA for count data (which we call ``Poisson SVD") when either sample size or feature dimension is limited. First, we adapt the iterative bootstrap method \citep{scholz2007bootstrap} to Poisson SVD to eliminate the bias of estimated loadings. We also prove that the estimator produced by this method achieves a second-order bias correction. Next, we consider SIMEX \citep{cook1994simulation}, a well-known simulation-based method for dealing with Poisson regression where covariates (loading estimates) follow the classical measurement error model, to further improve score estimates.  Lastly, we correct MLE bias in score estimates through the Firth correction \citep{firth1993bias}. Through extensive simulation studies and analysis of the million song dataset, we demonstrate the effectiveness of this bias correction procedure. 

The rest of the paper is organized as follows. Section~\ref{bg_ex} reviews Poisson SVD and presents three simulation examples to illustrate the source of bias of loadings and scores estimators for motivation. In Section~\ref{method}, we propose a bias correction procedure to improve estimated loadings and scores. Section~\ref{simulation} and Section~\ref{da} provide a comprehensive simulation study, and analysis of a subset of the million song dataset for recommendation of songs to demonstrate the effectiveness of the bias correction procedure. Finally, we conclude this work and discuss future research directions in Section~\ref{conclusion}.

\section{Background and Motivation\label{bg_ex}}
This section provides a review of Poisson SVD, a special case of exponential family PCA \citep{collins2001generalization} applied to count data. We also discuss three motivating examples to illustrate the bias of estimated loadings and scores by Poisson SVD through a simulation study. 
\subsection{Exponential family PCA on count data: Poisson SVD\label{bg}}
Exponential family PCA can be used to perform dimension reduction on multivariate count data. We refer to exponential family PCA applied to count data as Poisson SVD. Suppose $X = [x_{ij}]$ is an $n\times d$ count data matrix. We assume that each entry $x_{ij}$ is independently generated from a Poisson distribution with mean parameter $\lambda_{ij}$, and write concisely
\begin{equation*}
    X\sim \mbox{Poisson}(\Lambda), \mathrm{~where~}\Lambda = [\lambda_{ij}].
\end{equation*}
The distribution for $X$ can be represented using the natural parameter matrix $\Theta = [\theta_{ij}] = [\log(\lambda_{ij})]$. We posit that the underlying latent structure of $X$ comes from the natural parameter matrix being of low rank, say, $k$ with $k<d$:
\begin{equation}
\label{lowrank}
  \Theta =\bm{1}_n\bm{\mu}+ AV^\top,   
\end{equation}
where $\bm{1}_n$ denotes the $n$-dimensional vector of ones, $\bm{\mu}$ is a $d$-dimensional main effect vector, $A$ is an $n\times k$ matrix for component scores and $V$ is a $d\times k$ matrix for component loadings.

Poisson SVD estimates the main effects, loadings, and scores jointly by maximizing the likelihood of $X$:
$$
            \left(\hat{\bm{\mu}}, \hat{A},\hat{V}\right) = \argmax_{\bm{\mu}, A,V}~\ell(X;\bm{1}_n \bm{\mu}^\top + AV^\top),
$$
where $\ell(X;\Theta) = \sum_{i=1}^n\sum_{j=1}^d \ell(x_{ij}; \theta_{ij}) = \sum_{i=1}^n\sum_{j=1}^d \Bigl(x_{ij}\theta_{ij} - \exp({\theta_{ij}}) - \log(x_{ij}!) \Bigl )$ denotes the log-likelihood of count data $X$ under the Poisson distribution with a natural parameter matrix $\Theta$. When $\bm{\mu}$ is fixed, a simple way to perform this maximization is through the alternating algorithm, which alternatively optimizes one of $(A,V)$ at each step while setting the other as fixed. As $\bm{\mu}$ is treated as an offset, for simplicity,  $\bm{\mu}$ may be set to the logarithm of the column mean of $X$, or it can also be estimated through alternating steps. 

In the alternating steps, estimating component scores $A = [\bm{a}_1,\dots, \bm{a}_n]^\top$ can be viewed as fitting $n$ Poisson regression models with each row of data $\bm{x}_{i\cdot}$ as response, estimated main effects $\hat{\bm{\mu}}$ as offsets and estimated loadings $\hat{V}$ as covariates: 
$$
            \hat{\bm{a}}_i= \argmax_{\bm{a}\in\mathcal{R}^k}~\ell(\bm{x}_{i\cdot}; \hat{\bm{\mu}} + \hat{V}\bm{a}),~i = 1,\dots, n.
$$
Therefore, applying Poisson SVD to new data to obtain component scores is equivalent to fitting Poisson regression models. Similarly, estimating the loadings $V = [\bm{v}_1,\dots, \bm{v}_d]^\top$ can also be viewed as getting the coefficient estimates from $d$ Poisson regressions with each column of $\bm{x}_{\cdot j}$ as response, $\hat{\mu}_j$ as the offset and estimated scores $\hat{A}$ as covariates:
$$
            \hat{\bm{v}}_j= \argmax_{\bm{v}\in \mathcal{R}^k}~\ell(\bm{x}_{\cdot j}; \bm{1}_n\hat{\mu}_j + \hat{A}\bm{v}),~j = 1,\dots, d.
$$

Note that no constraint on loadings and scores is imposed in the algorithm. However, in order to identify loadings and scores parameters, introducing some constraints is necessary. For example, the ``identifiable up to sign" condition \citep{wang2021deviance} requires an orthonormal constraint on $V$ such that $V^\top V = I_{k}$ and a constraint on $A$ such that $A$ has pairwise orthogonal columns. Practically, we can adjust the estimates after we get $\hat{V}$ and $\hat{A}$ from the non-constrained algorithm to meet these requirements. Performing singular value decomposition (SVD) on $\hat{A}\hat{V}^\top$, for example, will naturally ensure that the estimated loadings and scores satisfy the constraints for being identifiable up to sign. 

\subsection{Motivating examples\label{ex}}
\begin{table}[t]
\begin{center}
\caption{Sample size ($n$), number of variables ($d$), and distributions used for generating main effects, loadings and scores for three examples. See Section~\ref{datagen} for details.}
\label{tab:my-table}
\begin{tabular}{cccccc}
\hline
\hline
 & $n$ & $d$ & Main effects & Loadings & Scores \\
\hline
Example 1 & 100 & 50 & $N(-2, 4)$ & $N(0, 1)$ & $N(0, 4)$ \\
Example 2 & 100 & 50 & $N(1, 4)$ & $N(2, 1)$ & $N(0, 4)$ \\
Example 3 & 100 & 200 & $N(-1, 4)$ & $N(0, 1)$ & $N(0, 4)$\\
\hline
\end{tabular}
\end{center}
\end{table}
When either sample size $n$ or feature dimension $d$ is not large enough, estimated loadings and scores tend to be biased. The following three examples illustrate different magnitudes of bias in the estimates. We simulate count data by specifying the natural parameter matrix in (\ref{lowrank}) with one component ($k=1$), $n = 100$ and $d = 50$ or $200$. Elements of $\Theta$ are generated from some probability distributions. Table~\ref{tab:my-table} displays the distributions for main effects, loadings, and scores that are used to generate three data examples. 
\begin{figure}[htbp]
    \centering
    \includegraphics[width = \textwidth]{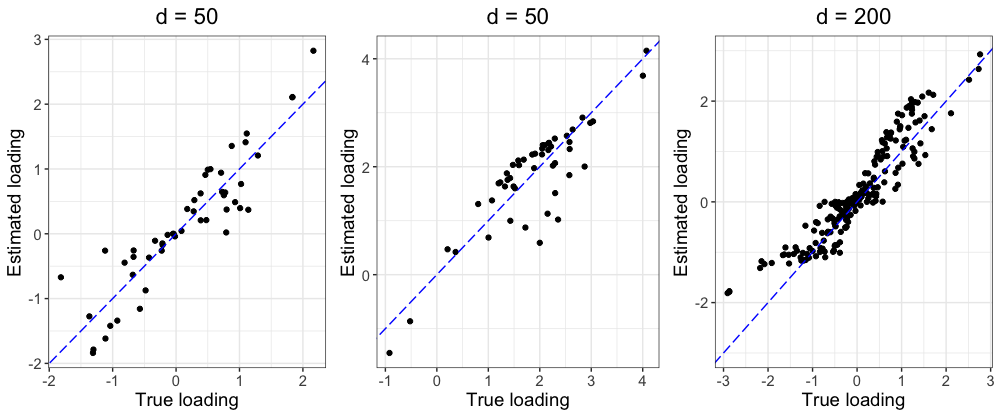}
    {(a) Example 1 \hspace{23mm}(b) Example 2 \hspace{23mm} (c) Example 3}
    \caption{Comparison of estimated loadings and true loadings for the three motivating examples. The blue dotted line denotes the 45-degree line indicating that the estimated loading equals the true loading.}
    \label{loadings_ex}
\end{figure}

First, we examine loading estimates. Figure~\ref{loadings_ex} shows the loading estimates versus true loading values. The dotted line is a 45-degree line that represents where the estimates perfectly align with the true values. Loading estimates may deviate from true loadings systematically as they may inflate or deflate the true values. The three examples highlight different bias patterns. In the first example (left), estimates appear to inflate or deflate the true loading values randomly, and the bias pattern has no clear association with the true loading values. In the second example (middle), a systematic pattern of inflation is present in the range where the true loading values are between 1 and 3 while other points are close to the 45-degree line. Different from the first two examples, deflation occurs in example 3 (right) on the left side where the true loading values are negative. These observations suggest the presence of a possible systematic bias in estimated loadings, and the magnitude of bias may vary among different examples.

\begin{figure}[htbp]
    \centering
    \includegraphics[width = \textwidth]{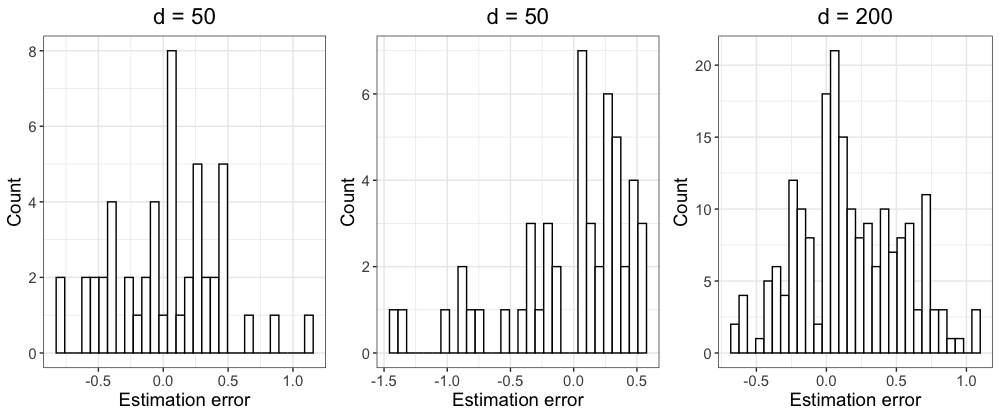}
    {(a) Example 1 \hspace{23mm}(b) Example 2 \hspace{23mm} (c) Example 3}
    \caption{Histograms of the estimation errors of loadings (estimated loading $-$ true loading) for the three motivating examples.}
    \label{error_ex}
\end{figure}

One way to measure the bias is to look at the estimation error of loadings (estimated loading $-$ true loading). Figure~\ref{error_ex} displays the histograms of the estimation errors of loadings for the three motivating examples.
If there exists a systematic bias, then the mean of estimation error should be far away from zero. We did one-sample t-tests to assess whether the mean equals zero. The first and second examples return $p$-values greater than 0.05 while the third example outputs an extremely small $p$-value much less than 0.05. This motivates us to further investigate how to reduce the systematic bias of the loading estimator in a finite sample case. 

\begin{figure}[htbp]
    \centering
    \includegraphics[width = \textwidth]{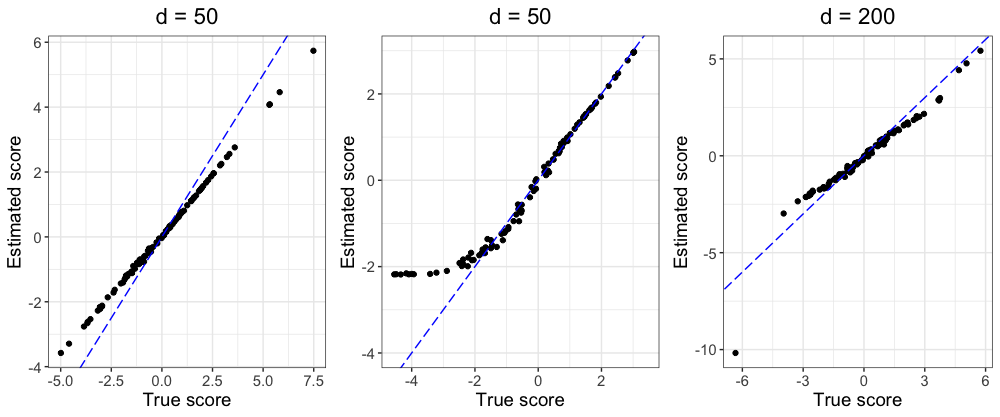}
    {(a) Example 1 \hspace{23mm}(b) Example 2 \hspace{23mm} (c) Example 3}
    \caption{Comparison of the estimated scores and true scores for the three motivating examples. The blue dotted line denotes the 45-degree line indicating that the estimated score equals the true score.}
    \label{score_ex}
\end{figure}

Furthermore, estimated scores are affected by estimated loadings substantially. As seen in Figure~\ref{score_ex}, scores estimates are prone to inflate or deflate the true scores due to the bias and errors in loading estimates. Besides, MLE bias is generally involved in this process, contributing to those poorly behaving score estimates as observed in exponential family PCA with binary data \citep{huang2023dissertation}.

\section{Methodology\label{method}}
Inspired by the observations from the motivating examples, we propose a bias correction procedure to reduce the bias along the estimation process in Poisson SVD. We begin by correcting the bias of the loadings estimator. After removing the bias, we consider incorporating a measurement error model to manage the random estimation error. MLE bias in the scores estimates is addressed in the final step. 

\subsection{Iterative Bootstrap Bias Correction}
Obtaining an unbiased estimator is often difficult, especially under complex models. Even the asymptotically unbiased maximum likelihood estimator has significant sample bias when the sample size is not large enough. Since the bias of an estimator usually cannot be derived analytically, simulation-based approaches such as the jackknife \citep{efron1982jackknife} or the bootstrap \citep{efron1994introduction} are broadly used to correct the bias. For example, vanilla bootstrap-based bias correction and its variants have been utilized in many statistical models, including generalized linear model with random effects \citep{kuk1995asymptotically}, dynamic panel model \citep{everaert2007bootstrap}, time series auto-regressive models \citep{tanizaki2004computational}, and generalized linear latent variable models \citep{guerrier2019simulation}.

The idea of bootstrap-based bias correction is to approximate the bias of a given estimator based on bootstrap samples and then correct the original estimator by removing the estimated bias. One natural extension is to perform bootstrapping iteratively. One iterative bootstrap method \citep{kuk1995asymptotically} for bias correction that has been well studied recently \citep{guerrier2019simulation, guerrier2020general, zhang2022flexible} is to keep updating the bias estimate by generating bootstrap samples based on the previous bias-corrected estimate. This method is proved to provide a computational solution to the optimization problem of the indirect inference \citep{gourieroux1993indirect} and can remove at least second-order bias. Another less known formulation of iterative bootstrap is proposed by \cite{scholz2007bootstrap}. Scholz argued that the bias-corrected estimator from the classical bootstrap still contains some bias and requires an additional iteration of bootstrapping to handle the new bias. The usage of this approach is only illustrated in two iterations. We decide to take this less understood Scholz's version of iterative bootstrapping, examine the method, and adapt it to Poisson SVD in order to correct the bias in estimated loadings. 

To be more concrete, we review the formulation of Scholz's iterative bootstrap and the idea behind it as follows. Suppose $\hat{\bm{\pi}}$ is a biased estimator of the true parameter $\bm{\pi}_0 \in \mathbb{R}^p$ from sample $Y$ of size $n$ such that
$$
E_{Y\sim f(\bm{\pi}_0)}[\hat{\bm{\pi}}] = \bm{\pi}_0 + \upsilon(\bm{\pi}_0), 
$$
where $f(\bm{\pi}_0)$ is the probability density function of $Y$ under the parameter $\bm{\pi}_0$ and $\upsilon(\cdot)$ denotes the bias as a function of $\bm{\pi}_0$. Suppose the classical bootstrap generates B samples, i.e., $Y^{(1)},\dots,Y^{(B)}$. Under the same estimation method, we have $\hat{\bm{\pi}}^{(1)},\dots, \hat{\bm{\pi}}^{(B)}$ for the corresponding $B$ bootstrap samples. Then we can have a bias-corrected estimator 
$\hat{\bm{\pi}}_{bs} = \hat{\bm{\pi}} - (\frac{1}{B}\sum_{b=1}^B\hat{\bm{\pi}}^{(b)} - \hat{\bm{\pi}})$. By the law of large numbers, as $B\rightarrow\infty$,
$$\hat{\bm{\pi}}_{bs} \overset{p}{\to} \hat{\bm{\pi}} - (\hat{\bm{\pi}} + \upsilon(\hat{\bm{\pi}}) - \hat{\bm{\pi}}) = \hat{\bm{\pi}} - \upsilon(\hat{\bm{\pi}}) \coloneqq \tilde{\bm{\pi}}_{bs}.
$$
Unfortunately, even under a large number of bootstrap samples, $\tilde{\bm{\pi}}_{bs}$ is highly likely to be a biased estimator as well since 
$E_{Y\sim f(\bm{\pi}_0)}(\tilde{\bm{\pi}}_{bs}) = \bm{\pi}_0 + \upsilon(\bm{\pi}_0) - E_{Y\sim f(\bm{\pi}_0)}[\upsilon(\hat{\bm{\pi}})]$. As a result, one more round of bootstrap is taken into account to handle the new bias $\upsilon_1(\bm{\pi}_0) = \upsilon(\bm{\pi}_0)-E_{Y\sim f(\bm{\pi}_0)}[\upsilon(\hat{\bm{\pi}})]$ brought by $\tilde{\bm{\pi}}_{bs}$. In the second iteration, we generate $C$ bootstrap samples for each $\hat{\bm{\pi}}^{(b)}$ denoted by $Y^{(bc)}$ for $c = 1,\dots,C$. Let $\hat{\bm{\pi}}^{(bc)}$ be the estimate from $Y^{(bc)}$. The desired iterative bootstrap bias-corrected estimator is then of the form
\begin{align}
\label{ib_eq}
    \tilde{\bm{\pi}}_{ib} &= \tilde{\bm{\pi}}_{bs} - \upsilon_1(\hat{\bm{\pi}})\nonumber\\
    & = \hat{\bm{\pi}} - \upsilon(\hat{\bm{\pi}}) - \{\upsilon(\hat{\bm{\pi}}) - E_{Y\sim f(\hat{\bm{\pi}})}[\upsilon(\hat{\bm{\pi}})] \}\nonumber\\
    & = \hat{\bm{\pi}} - 2\upsilon(\hat{\bm{\pi}}) + E_{Y\sim f(\hat{\bm{\pi}})}[\upsilon(\hat{\bm{\pi}})]\nonumber\\
    & = \{\hat{\bm{\pi}} + \upsilon(\hat{\bm{\pi}}) + E_{Y\sim f(\hat{\bm{\pi}})}[\upsilon(\hat{\bm{\pi}})] \} - 3\upsilon(\hat{\bm{\pi}}).
\end{align}
The term in the curly brackets of Equation~\ref{ib_eq} can be estimated using the estimates from the bootstrap samples of the second round since by the law of large numbers, for each $b$, as $C \rightarrow \infty$,
$$
\frac{1}{C}\sum_{c=1}^C\hat{\bm{\pi}}^{(bc)}\overset{p}{\to} \hat{\bm{\pi}}^{(b)} + \upsilon(\hat{\bm{\pi}}^{(b)}),$$
which implies that as $B\rightarrow \infty$ and $C \rightarrow \infty$,
$$
\frac{1}{B}\sum_{b=1}^B\frac{1}{C}\sum_{c=1}^C\hat{\bm{\pi}}^{(bc)}\overset{p}{\to} \hat{\bm{\pi}} + \upsilon(\hat{\bm{\pi}}) + E_{Y\sim f(\hat{\bm{\pi}})}[\upsilon(\hat{\bm{\pi}})].
$$
Therefore, the final bias-corrected estimator is
\begin{align}
\label{ib_est}
\hat{\bm{\pi}}_{ib} = 3\hat{\bm{\pi}} - \frac{3}{B}\sum_{b=1}^B\hat{\bm{\pi}}^{(b)} + \frac{1}{B}\sum_{b=1}^B\frac{1}{C}\sum_{c=1}^C \hat{\bm{\pi}}^{(bc)}.
\end{align}

We further investigate a statistical property of this estimator. Following \cite{guerrier2019simulation}, we start with the following assumption about the form of the bias $\upsilon(\cdot)$ of the initial estimator $\hat{\bm{\pi}}$ that it is composed of two terms: one constant term and the other term related to the true parameter.
\begin{assumption}
\label{ib_assumption}
The bias vector $\upsilon(\bm{\pi})$  can be expressed as a sum of two parts:
$$
\upsilon(\bm{\pi}) =\bm{c}(n)+\bm{t}(\boldsymbol{\pi}, n).
$$
Here, $\bm{c}(n)$ is a constant with respect to $\bm{\pi}$, bounded, converges to a constant $\mathbf{c}$ when $n \rightarrow$ $\infty$, and
$$
\mathbf{c}(n)=\mathcal{O}\left(n^{-\beta}\right), \quad \beta \geq 0.
$$
For $\bm{t}(\boldsymbol{\pi}, n)$, there exists $r_{i, j} \in \mathcal{R}, s_{k, l, j} \in \mathcal{R}$, for $i, j, k, l=1, \ldots, p$ such that
$$
\bm{t}(\boldsymbol{\pi}, n)=\left[\sum_{i=1}^p r_{i, j} \frac{\pi_i}{n}+\sum_{k=1}^p \sum_{l=1}^p s_{k, l, j} \frac{\pi_k \pi_l}{n^2}+\mathcal{O}\left(n^{-3}\right)\right]_{j=1, \ldots, p}.
$$
\end{assumption}
Under this assumption, we establish the convergence rate of the bias of Scholz's iterative bootstrap bias-corrected estimator and state it in the following theorem.
\begin{theorem} Under assumption 1, Scholz's iterative bootstrap bias-corrected estimator $\hat{\bm{\pi}}_{ib}$ satisfies
\label{ib_thm}
$$
E[\hat{\bm{\pi}}_{ib}] = \bm{\pi}_0 + \mathcal{O}(n^{-2}).
$$
\end{theorem}

Theorem~\ref{ib_thm} states that Scholz's bias correction method produces an estimator that can achieve a second-order bias correction. The classical bootstrap bias-corrected estimator, as stated in \cite{guerrier2019simulation}, can only achieve at best a second-order correction. So, Scholz's iterative bootstrap method could be better than the classical bootstrap method in terms of the convergence rate of the bias. The proof of Theorem~\ref{ib_thm} is provided in Appendix \ref{ib_proof}.

We choose Scholz's iterative bootstrap method to remove bias in loadings. However, 
applying the method to Poisson SVD requires additional decisions, some of which are more specific to the method. The first question is to choose between nonparametric bootstrap and parametric bootstrap. Note that nonparametric bootstrap does not require distribution assumptions, and applying nonparametric bootstrap to count data is equivalent to resampling the count data. Consequently, the effective sample size becomes smaller and this leads to even worse loadings estimates. Therefore, parametric bootstrap is more suitable for Poisson SVD where we already make a distribution assumption on the count data. Second, Poisson SVD actually has two sets of parameters of interest, i.e., loadings and scores. However, we only focus on the bias of loadings thus scores can be treated as nuisance parameters. After generating the bootstrap samples, we still jointly estimate loadings and scores by maximizing the likelihood. We only use estimated loadings ($\hat{V}$), and estimated mean parameter ($\hat{\Lambda}$) for generation of bootstrap samples in the second iteration. Finally, model identifiability is an implicit assumption for applying bootstrap. Loadings from Poisson SVD can be identifiable up to sign by imposing some constraints \citep{wang2021deviance}. To determine the polarity of loading vectors, we choose the sign of the estimated loadings as whichever one brings them closer to the original estimated loadings. 

Algorithm~\ref{alg:ib} summarizes the steps of the iterative bootstrap bias correction for Poisson SVD. Empirically, we find that instead of generating $B\times C$ samples by classical bootstrap, generating B samples in the first iteration and for each generated sample, generating C samples in the second iteration produces larger bias reduction and leads to better loading estimates (see Appendix~\ref{bootstrap}), which is consistent with the result in Theorem \ref{ib_thm}.

\begin{algorithm}[htbp]
\begin{algorithmic}[1]
\REQUIRE  Original estimated loadings ($\hat{V}$), mean parameter ($\hat{\Lambda}$), main effects ($\hat{\bm{\mu}}$), number of components ($k$), number of bootstrap samples in the first iteration (B), number of bootstrap samples in the second iteration (C) 
\ENSURE Bias-corrected loadings estimates ($\tilde{V}$) 
\FOR{$i=1$ to $B$}
\STATE Draw a bootstrap sample $X^{(i)}\sim \mbox{Poisson}(\hat{\Lambda})$
\STATE Apply Poisson SVD to $X^{(i)}$ and obtain $\hat{V}^{(i)}$ and $\hat{A}^{(i)}$
\STATE ($\tilde{V}^{(i)}, \tilde{A}^{(i)}$) = SVD($\hat{A}^{(i)}(\hat{V}^{(i)})^\top$) with rank $k$ and tr$((\hat{V}^{(i)})^\top \hat{V})>0$
\STATE $\hat{\Lambda}^{(i)} = \exp(\bm{1}_n\hat{\bm{\mu}}^\top +\tilde{A}^{(i)}(\tilde{V}^{(i)})^\top)$
\FOR{$j=1$ to $C$}
\STATE Draw a bootstrap sample $X^{(ij)}\sim Poisson(\hat{\Lambda}^{(i)})$
\STATE Apply Poisson SVD to $X^{(ij)}$ and obtain $\hat{V}^{(ij)}$ and $\hat{A}^{(ij)}$
\STATE $(\tilde{V}^{(ij)}, \tilde{A}^{(ij)})$ = SVD($\hat{A}^{(ij)}(\hat{V}^{(ij)})^\top$) with rank $k$ and tr$((\hat{V}^{(ij)})^\top \hat{V})>0$
\ENDFOR
\ENDFOR
\STATE $\tilde{V} = 3\hat{V} - \frac{3}{B}\sum_{b=1}^B\tilde{V}^{(b)} + \frac{1}{B}\sum_{b=1}^B\frac{1}{C}\sum_{c=1}^C \tilde{V}^{(bc)}$
\end{algorithmic}
\caption{Iterative bootstrap bias correction for Poisson SVD }
\label{alg:ib}
\end{algorithm}

\subsection{Measurement Error Model}
\begin{figure}[htbp]
    \centering
    \includegraphics[scale=0.7]{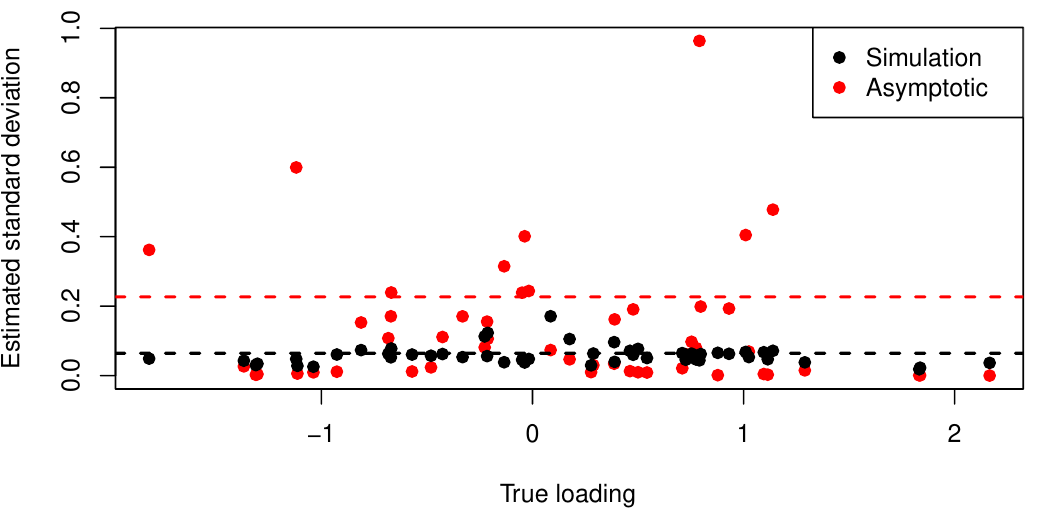}
    \caption{Estimated standard deviations of loadings versus  true loadings for the first motivating example in Section~\ref{ex}.  The red points represent estimates from the asymptotic distributions based on the same data while the black points denote sample standard deviations from 50 random samples generated by true parameters. The red and black dashed lines indicate the average of standard deviations estimated in these two ways.}
    \label{sd}
\end{figure}

Even after the systematic bias of the loadings estimator has been removed by the first step,  the estimated loadings may still deviate from the true loadings due to sampling errors. Therefore, the estimated loadings $\hat{V}$ can be further treated as contaminated observations of the true loadings $V^*$. The classical additive measurement error model \citep{grace2021handbook} may be suitable to describe the relationship between estimated loadings (observed) and true loadings (unobserved),
\begin{align}
\label{measurement_error}
\hat{v}_{ij} = v^*_{ij} +\epsilon_{ij}, ~\epsilon_{ij} \sim N(0,\sigma^2), \mathrm{~for~}i = 1,\dots,d,~j = 1,\dots,k.
\end{align}
Here, $\epsilon_{ij}$ is called the measurement error, and this error is assumed to be normally distributed with zero mean and unknown variance $\sigma^2$. In this way, score estimates can be further improved by utilizing measurement error models for generalized linear models. Before jumping to methods for Poisson regression models with errors-in-variables, two important questions left are whether the assumption of the same measurement error variance for all loading elements is reasonable, and how to estimate the measurement error variance accurately.\\
We did a sanity check on the same variance assumption through simulation. Given true parameters, we can simulate multiple count data matrices based on Poisson distributions with the true parameters. For each set of simulated data, we apply Poisson SVD to get loadings estimates. Letting $\sigma^2_{ij}$ be the variance of $\hat{V}_{ij}$ for $i= 1,\dots,d$ and $j = 1,\dots, k$, we can estimate $\sigma^2_{ij}$ by elementwise variance of those loadings matrices.
If $\sigma_{ij}$'s are similar and have smaller deviation, this indicates that $\sigma_{ij}$ can be simplified to $\sigma$ as in model (\ref{measurement_error}). Besides, since estimation of loadings is related to Poisson regression, we can also estimate $\sigma^2_{ij}$ using the asymptotic variance for loadings. For each row of the loadings matrix,
$$
\hat{\bm{v}}_i \overset{d}{\approx} N_k\left(\bm{v}_i,(\hat{A}^\top W_a\hat{A})^{-1}\right),~i = 1,\dots, d,
$$
where $W_a  = \mathrm{diag}(w_1,\dots,w_n)$ and $w_m = \exp(\hat{\mu}_m + \hat{\bm{a}}_m^\top \hat{\bm{v}}_i),~m = 1,\dots,n$. We take the diagonal elements of the asymptotic covariance matrix as a variance estimate of each element in $\hat{\bm{v}}_i$. 
\\
Here we show the standard deviation of loadings for the first simulation example in Section~\ref{ex} for illustration. We generate 50 random samples given the true parameters of $\Theta$. We calculate the standard deviations for each element of estimated loadings and plot them as black points in Figure~\ref{sd}. There is not much difference among the standard deviations, which justifies the equal variance assumption in this example.   Note that some standard deviations estimated from the asymptotic distributions (red points) are close to zero while others are much larger than their sample counterparts (black points). This is because when $\hat{\bm{v}}_i$ is large in the absolute value, $W_a$ tends to have large diagonal values, resulting in nearly zero asymptotic variances and vice versa for small $\hat{\bm{v}}_i$. A naive estimate of $\sigma^2$ is the  mean variance from the asymptotic distributions. In this example, $\hat{\sigma}$ from the asymptotic distribution is larger than $\hat{\sigma}$ from the simulation. A larger $\hat{\sigma}$ implies that we assume a larger measurement error in loadings, which, in turn, leads to more substantial changes in the score estimates.

There exist many methods for Poisson regression where covariates have measurement error, and they can be utilized to incorporate the error in estimating loadings to achieve better scores estimates. 
One category of those methods is designed to modify the likelihood. \cite{carroll1984errors} integrated the distribution of true covariates to obtain the likelihood of data. Regression calibration \citep{carroll1990approximate, armstrong1985measurement} uses the best approximation of true covariates by taking the measurement error into account. \cite{stefanski1987conditional} proposed the conditional score approach, treated unobserved true covariates as unknown parameters, and found the sufficient statistics for the true covariates. Also, score functions could be corrected such that the expectation is identical to the score functions without measurement errors \citep{stefanski1989unbiased}. Another widely-used approach is based on simulation: SIMulation EXtrapolation (SIMEX) \citep{cook1994simulation}. SIMEX attempts to establish the relationship between naive estimates (estimated under the error-free setting) and the size of measurement error variance through simulated samples. Then, one extrapolates back to the setting with no measurement error.  We choose SIMEX in our simulation study to evaluate whether considering measurement error could bring better score estimates since this method is easy to implement while still guaranteeing desired statistical properties. 

\subsection{MLE Bias Reduction}
To further improve score estimates, we attempt to remove the first-order asymptotic MLE bias. \cite{firth1993bias} proposed a well-known MLE bias correction method for generalized linear model.  The idea is to correct the bias implicitly through adjusting the score equation, which is equivalent to adding a Jeffreys invariant prior as a penalty function to the likelihood function \citep{kosmidis2009bias}. The score estimate by Firth's correction is
$$
            \hat{\bm{a}}_{Firth}= \argmax_{\bm{a}\in \mathcal{R}^k}~\ell(\bm{x}_{i\cdot}; \hat{\bm{\mu}} + \hat{V}\bm{a}) + \frac{1}{2}\log|I(\bm{a})|,
$$
where $I(\bm{a}) = \hat{V}^\top W_v \hat{V}$ is the Fisher information matrix, and $W_v  = \mathrm{diag}(\exp(\hat{\bm{\mu}}+\hat{V}\bm{a}))$. 
This is an extremely useful method to deal with sparse data and unbounded MLEs in Poisson regression. For example, when responses are all zeros, fitting a Poisson regression leads to an infinite MLE, but applying Firth's correction prevents such situations and outputs reasonable MLEs. 
  
\section{Simulation Study\label{simulation}}
In this section, we demonstrate the effectiveness of the proposed bias correction procedure on simulated count data matrices. We examine the effects of iterative bootstrap bias correction for loadings, accounting for measurement error in loadings and MLE bias correction for scores, separately. 
\subsection{Simulation Setup\label{datagen}}
For a simulated dataset, we start with a natural parameter matrix $\Theta = [\log(\lambda_{ij})]$ of the form in (\ref{lowrank}) and each entry of the count data matrix $X_{ij}$ is independently sampled from a Poisson distribution with mean $\lambda_{ij}$. We consider different setups for parameters to examine whether each bias correction step could improve the estimates of Poisson SVD: (1) $n = 100,~d\in\{50, 100, 200\}$; (2) $k \in\{1, 2\}$, and the setting with $k = 2$ is specifically used for evaluating the scenario in which $k$ is misspecified; (3) Each main effect element $\mu_i\sim N(c,4)$ with $c\in\{-3,-2,-1,0,1\}$, is assumed to be known in order to highlight the improvements in estimated loadings and scores brought by the bias correction procedure. Therefore, $\bm{\mu}$ can be treated as offsets, and the center of offsets distribution $c$ is associated with the sparsity level of the data; (4) Each entry of loadings matrix $v_{ij} \sim N(0,1)$; (5) When $k = 1$, we generate individual scores  $a_i\sim N(0,4)$. When $k = 2$, scores for the first component come from $N(0,4)$ and scores for the second component are from the standard normal distribution. For each combination of the above parameter settings, we simulate 50 datasets.

To evaluate the quality of loading estimates, we use the angle between estimated loadings $\hat{V}$ and true loadings $V^*$, 
$\frac{\mathrm{tr}(\hat{V}^\top V^*)}{\sqrt{\|\hat{V}\|^2_F\cdot\|V^*\|^2_F}}$ as an evaluation metric. A larger angle indicates a larger discrepancy between estimated loadings and true loadings. The quality of scores estimates $\hat{A}$ is assessed through the root mean square error (RMSE), i.e., $\sqrt{\frac{1}{nk}\|\hat{A}-A^*\|_F^2}$. Similarly, we could use RMSE to measure the quality of natural parameter estimates. To show the improvement of estimates brought by the bias correction procedure quantitatively, we define RMSE reduction percentage as $(\mathrm{RMSE_{before}  - RMSE_{after}}) / \mathrm{RMSE_{before}} \times 100\%$, where $\mathrm{RMSE_{before}}$ is the RMSE of estimates before bias correction and $\mathrm{RMSE_{after}}$ is the RMSE of bias-corrected estimates. This metric tells how much reduction in the estimation error the bias correction method brings.

\subsection{Iterative Bootstrap Bias Correction in Loadings \label{IB}}
\begin{figure}[tbp]
    \centering
    \includegraphics[width = \textwidth]{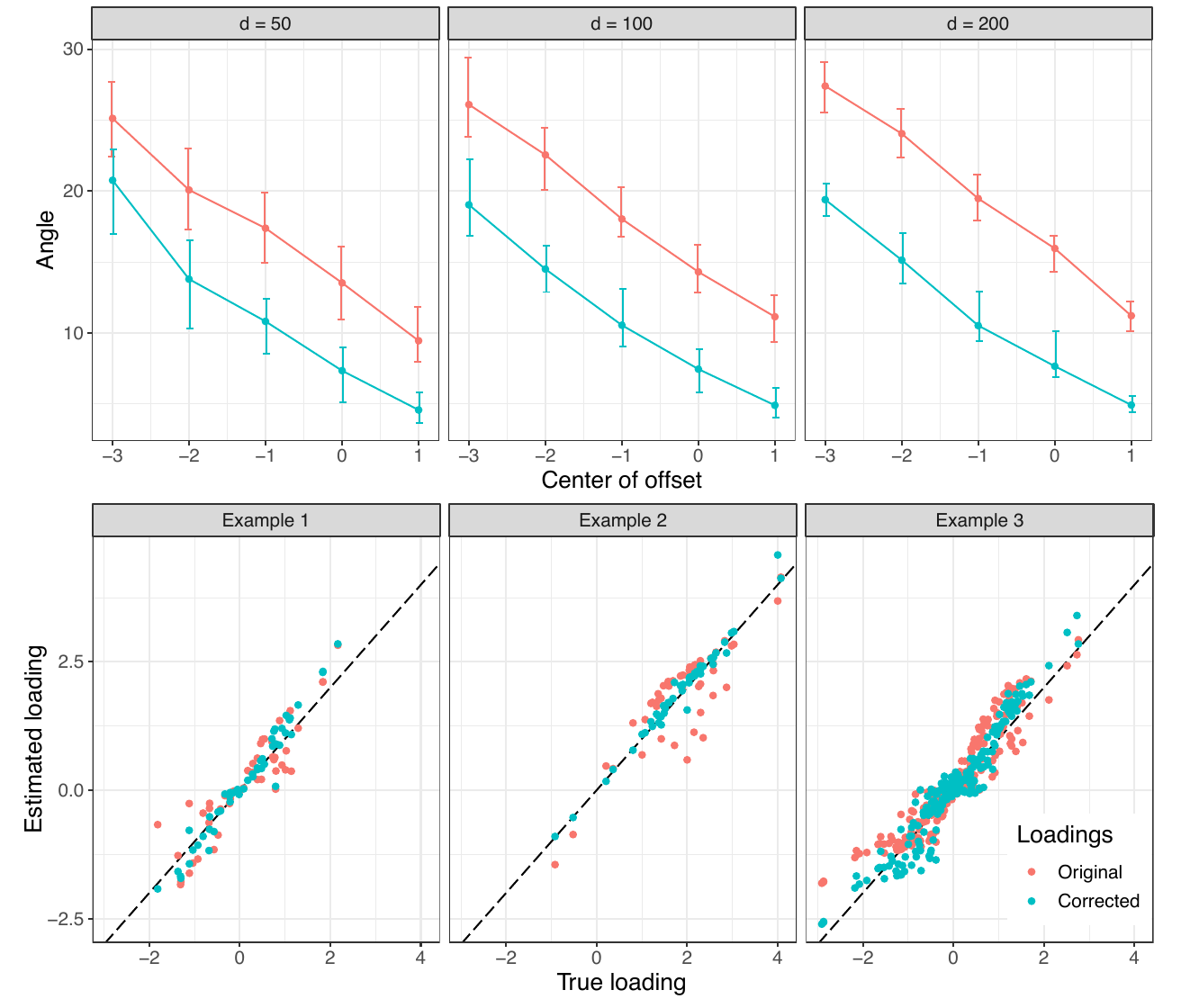}
    \caption{Top panel: Comparison of angles between the true loadings and estimated loadings for original estimates and iterative bootstrap bias-corrected estimates. Experiments are done  under $k=1$, different mean of offsets and $d$. The point in the middle of each bar denotes the median of replicates, and the bar ranges from the first to third quartiles. Bottom panel: Comparison of the true loadings and estimated loadings (original or bias-corrected). The dotted lines indicate that the estimated loading is equal to the true loading. }
    \label{k=1_angle}
\end{figure}

We first investigate the effect of iterative bootstrap for correction of bias in loadings. The upper panel of Figure~\ref{k=1_angle} shows the comparison of angles between the estimated loadings and true loadings under the scenario where there is one true component ($k = 1$) and the number of components is correctly specified. As data becomes more sparse with smaller main effects and $d$ gets larger, loading estimates from Poisson SVD tend to degrade as indicated by increasing angles. In general, performing bias correction on loadings with iterative bootstrap brings substantial reductions in angles regardless of the sparsity level of data and feature dimension. To show how the decrease in angle is associated with better loading estimates, the lower panel of Figure~\ref{k=1_angle} revisits the three simulation examples in Section~\ref{ex}. Bias-corrected loadings have a smaller estimation error as expected, which makes the points closer to the 45-degree line.

\begin{figure}[thbp]
    \centering
    \includegraphics[width = \textwidth]{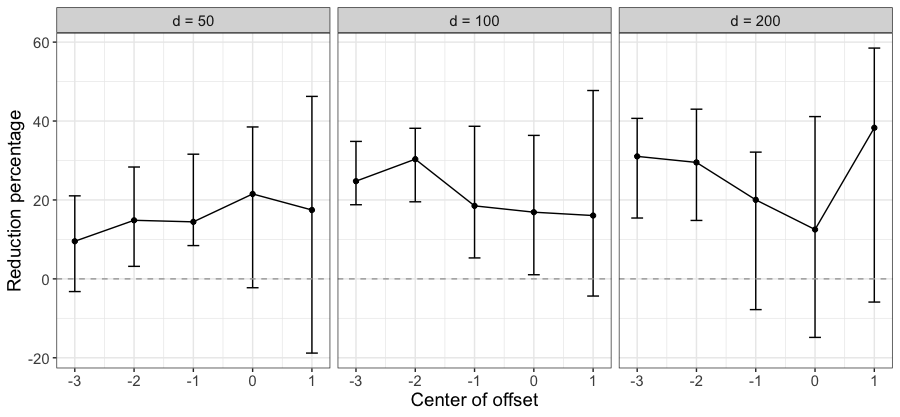}
    \caption{Percentage of reduction in RMSE of scores under $k = 1$, different center of offsets and data dimension $d$.}
    \label{k=1_score}
\end{figure}

In addition, we find that improvements in loadings estimates by iterative bootstrap lead to improvements in scores estimates as shown in Figure~\ref{k=1_score} in terms of RMSE reduction percentages. The RMSE of scores for the original loadings estimates is generally larger than that for the bias-corrected loadings. However, it is worth mentioning that the reduction percentages under some scenarios are below zero, which indicates that RMSE of scores has increased after correction on loadings. This may be attributed to that: (1) for some simulated cases, the angle actually becomes larger, indicating that the loadings estimates become worse, which leads to worse scores estimates; (2) there are other potential biases involved in the score estimation process, namely, MLE bias or numerical issues (e.g., number of maximum iterations has reached when fitting Poisson regression models). 

In practice, the true rank $k$ is usually unknown and the choice of number of components relies on selection criteria that we use such as cumulative deviance explained or BIC. Therefore, it is very likely that the number of components is misspecified. To explore how model misspecification affects the effectiveness of iterative bootstrap, we consider the scenario where there exist two true components but we fit a low-rank model with $k = 1, 3, 4$. In this case, the quality of loadings and scores estimates cannot be assessed directly. Hence, we evaluate estimates in terms of the natural parameters. We only present the results for $d = 50$ in Figure~\ref{k=2} as $d = 100$ or $200$ produces similar results (see Appendix~\ref{k=2_more}). 
\begin{figure}[htbp]
    \centering
    \includegraphics[width = \textwidth]{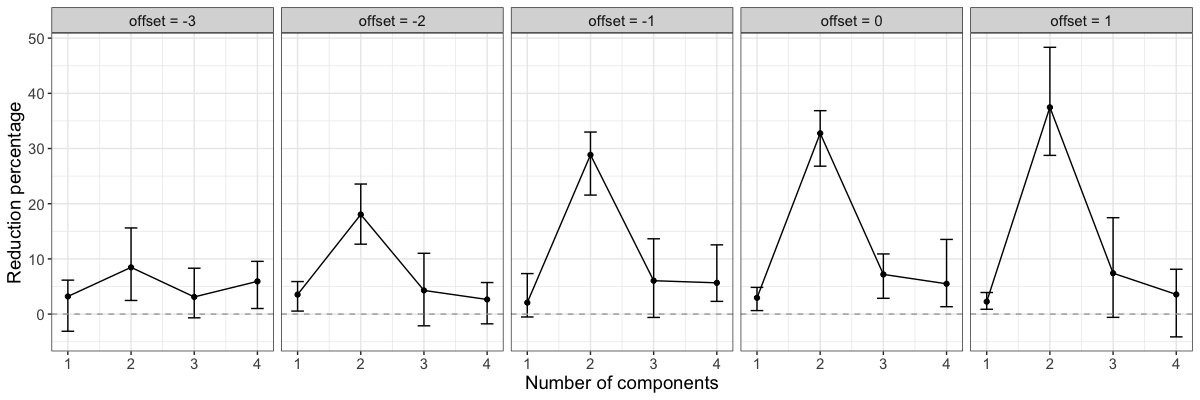}
    \caption{Percentage of reduction in RMSE of natural parameters when the true rank $k$ is misspecified. Experiments are done with two true components, $d = 50$, and different center of offsets.}
    \label{k=2}
\end{figure}
Iterative bootstrap yields a 5--10\% reduction in RMSE when the rank is misspecified. However, the extent of reduction is less compared to the correctly specified case.

\subsection{Accounting for Measurement Errors in Loadings}
After reducing the bias in loadings estimates by the iterative bootstrap method, we further apply SIMEX to deal with the remaining sampling error in loadings for estimation of scores and assess whether it may help improve score estimates. To apply SIMEX, we need an accurate estimate of the variance of random measurement error $\sigma^2$. We take the average of the variance of individual loading elements from asymptotic distributions as an estimate of $\sigma^2$. The square root of the variance estimate is then used as the standard deviation estimate in SIMEX. 
\begin{figure}[htbp]
    \centering
    \includegraphics[width = 0.7\textwidth]{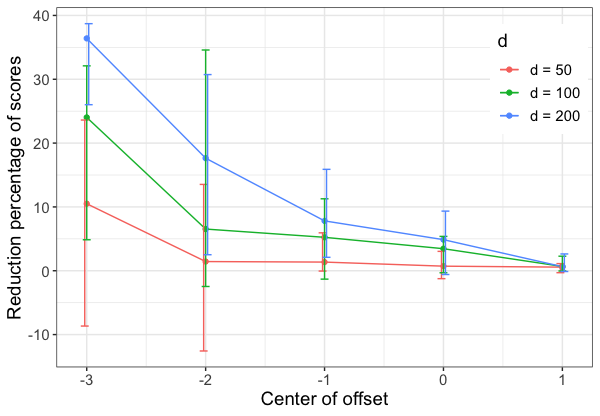}
    \caption{Percentage of reduction in RMSE of scores when comparing scores estimated with iterative bootstrap bias-corrected loadings and those with further application of SIMEX. Experiments are done with a single true component ($k=1$), and the number of components is correctly specified.}
    \label{SIMEX_better}
\end{figure}
Figure~\ref{SIMEX_better} shows the percentage of reduction in score estimates under different data dimensions $d$ and data sparsity levels. When data becomes less sparse with larger main effects, SIMEX provides less correction in scores estimates as the reduction percentage is almost less than 5\%. 
SIMEX also brings more improvements in scores under a larger $d$. After bias correction in loadings, the angles between the estimated loadings and true loadings are similar for different $d$. Since a larger $d$ indicates that more observations are used to estimate a score, the RMSEs of scores estimated using bias-corrected loadings are slightly better under $d = 200$. Moreover, we find that the estimated standard deviation for the measurement error tends to be larger on average for a larger $d$. This indicates that accounting for appropriate measurement errors may lead to better estimation results.

\subsection{MLE Bias Correction in Scores}
After diving into the iterative bootstrap method and measurement error issue in the previous sections, we now focus on the effect of MLE bias correction.
We find that MLE bias correction brings more effective changes to extreme MLEs.
In Poisson regression, the maximum likelihood estimates could be extreme when covariates can predict the response almost perfectly. A special example in Poisson regression is when the responses are all zeros, possibly producing unbounded MLEs. Since estimating scores using Poisson SVD is intrinsically connected to Poisson regression, such problems can happen with a count data matrix when there exist rows where each row element is zero.

\begin{figure}[htbp]
    \centering
    \includegraphics[width = 0.75\textwidth]{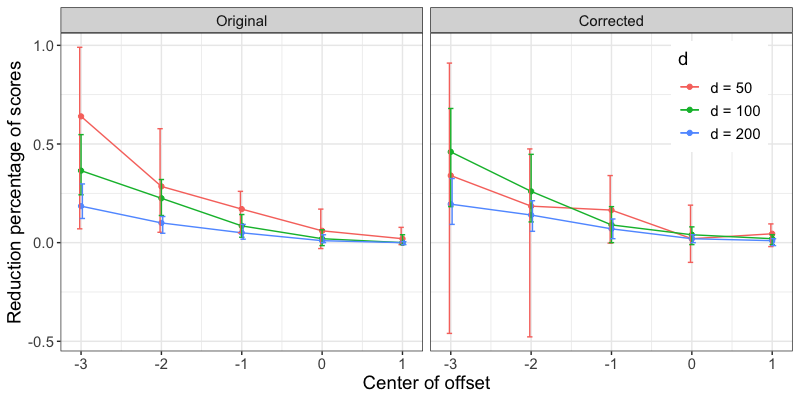}
    \caption{Percentage of reduction in RMSE of scores using the same simulation settings as in Figure~\ref{k=1_angle}.  The left panel presents Firth's MLE bias correction for scores estimated using original loadings estimates while the right panel is for those estimated using the bias-corrected loadings. For both settings, MLE bias correction provides trivial benefits.}
    \label{mle_bad}
\end{figure}

\begin{figure}[htbp]
    \centering
    \includegraphics[width = 0.75\textwidth]{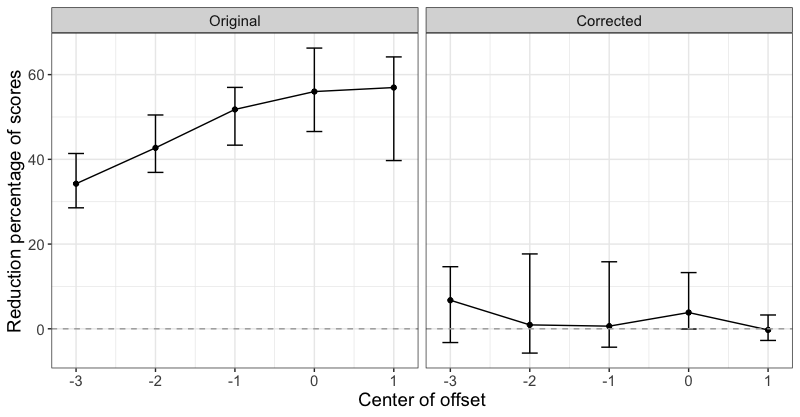}
    \caption{Percentage of reduction in RMSE of scores when the true rank $k = 1$, true loadings are from $N(-2,1)$, $d = 50$. The left panel presents Firth's MLE bias correction for scores estimated using original loadings estimates while the right panel is for those estimated using the bias-corrected loadings.}
    \label{mle_good}
\end{figure}

We illustrate how this affects MLE bias correction in two simulation settings. 
Figure~\ref{mle_bad} presents the reduction percentage in RMSE of scores under the same simulation scenarios as in Section~\ref{IB} with $k = 1$, where extreme MLEs are less likely to occur. 
The MLE bias correction step results in a trivial change in RMSE of scores as the reduction percentage is less than 1\%. Besides, when $d$ increases, the scale of MLE bias reduces, thus resulting in less effect on score estimates. Therefore, we only display $d = 50$ case in Figure~\ref{mle_good} for count data with this extreme MLE issue. By generating loadings from $N(-2,1)$,  loadings are more likely to have the same negative signs. With negative main effects and a positive score, there is a higher chance that we generate an observation with zeros for all features, which leads to an extreme MLE if we fit Poisson regression on this observation. Huge improvements can be observed in Figure~\ref{mle_good} for MLE bias-corrected scores in Poisson SVD using the original loadings estimates. When scores are estimated with bias-corrected loadings using the iterative bootstrap method, correcting MLE bias can further improve score estimates. 

Moreover, under this extreme MLE situation in Figure~\ref{mle_good}, we find that SIMEX only produces a reduction of less than 1\%  when $d= 50$ or $100$ and less than 4\% when $d = 200$. These results indicate that SIMEX does not significantly improve the accuracy of the score estimates.
Therefore, if the data are sparse and contain rows with all zeros, we recommend MLE bias correction in scores without SIMEX. Otherwise, adjusting scores through SIMEX without MLE bias correction will be sufficient.

\section{Data Analysis\label{da}}
In this section, we analyze a subset of the million song dataset (MSD) \citep{bertin2011million} to show the benefit of the proposed bias correction procedure, including the iterative bootstrap method, SIMEX or MLE bias correction.  The subset from MSD contains the listening history of around 75 thousand users on ten thousand songs. This data can be arranged as a sparse count matrix, where the users are in rows and the songs are taken as column variables. Each entry of this matrix represents the listening frequency of a specific song from a specific user. Following \cite{landgraf2020generalized}, we also consider the task of recommending new songs to users. A relatively new user may have less exposure to songs they might be interested in. Assuming that there exists a latent low-dimensional structure generating the associations between songs and users, we wish to capture such a structure by Poisson SVD and use our estimate to make recommendations to users. More specifically, we aim to obtain the latent component loadings and main effects by applying Poisson SVD to training data. Then, when new users come with an incomplete listening history, we can estimate ${\lambda}_{ij}$ from the low-dimensional representation of the listening counts for the users. A higher $\hat{\lambda}_{ij}$ indicates a greater interest in the $j$th song by the $i$th user. For example, a zero count means that the user has never listened to the song before, but if it comes with a larger $\hat{\lambda}$, this can be seen as a strong signal that this song might be of potential interest to the user. 

Next, we describe how training and test sets are specified for the task and the evaluation metric used.
Considering the limited computational resource, we choose a smaller subset of the data to work on. To take different sparsity levels of data into account, we start with drawing two subsets of popular songs that have been listened to by at least 500 users and 300 users, respectively. This leads to subsets of 486 songs and 1278 songs. We further randomly pick 50 songs and 125 top fans of those songs, and then randomly split 100 of the fans into a training set and 25 into a test set. To mimic ``new" songs for test users, we force five songs with positive counts to zero counts for each test user. If the model predicts well, these five songs should have higher $\hat{\lambda}$ compared to those songs with true zero counts. Since we have true positives (forced zero counts) and true negatives (true zero counts), we can use the area under the curve (AUC) of binary predictions (positive or negative) from a fitted model for users in the test set to evaluate the model performance. 

\begin{figure}[htbp]
    \centering
    \includegraphics[width = 0.9\textwidth]{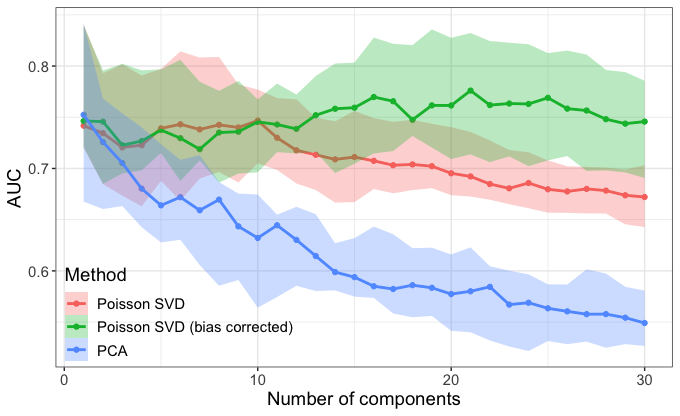}
    \caption{Comparison of Poisson SVD, debiased Poisson SVD, and standard PCA in terms of AUC values under different numbers of components when we select 50 songs from 1278 popular songs. We randomly select 50 songs for 20 replications. The lines represent the median AUC values. The color bands span from the first quartile to the third quartile of AUC.}
    \label{1278 songs}
\end{figure}
\begin{figure}[htbp]
    \centering
    \includegraphics[width = 0.9\textwidth]{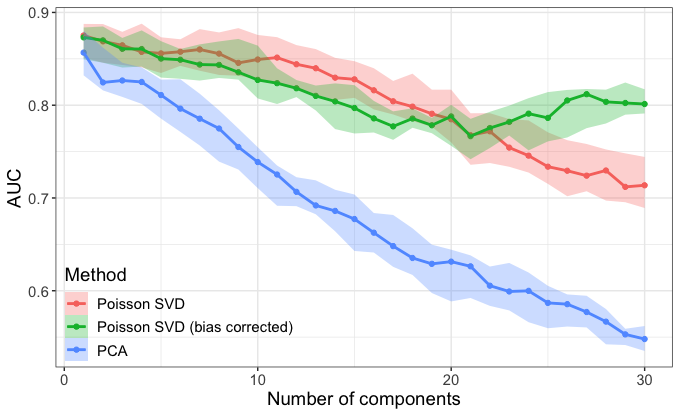}
     \caption{Comparison of Poisson SVD, debiased Poisson SVD, and standard PCA in terms of AUC under different numbers of components when we select 50 songs from 486 popular songs.}
     \label{486 songs}
\end{figure}

We compare Poisson SVD, Poisson SVD with the proposed bias correction procedure, and standard PCA in terms of the AUC under different numbers of principal components varying from 1 to 30 in Figure~\ref{1278 songs} and Figure~\ref{486 songs}. Standard PCA on the original count data has the poorest performance and its AUC keeps decreasing as the number of components increases. Note that Figure~\ref{1278 songs} corresponds to a highly sparse data matrix with around 80\% of zeros, and Figure~\ref{486 songs} corresponds to a moderately sparse data matrix with nearly 60\% of zeros. For both sparsity levels, the bias correction procedure can provide better AUC than the original estimates when $k$ is large while maintaining a similar performance as the original estimates when $k$ is small. Under the high level of sparsity, the bias correction procedure starts to show advantages when $k > 10$. However, with less sparse data, improvements in AUC happen at a much larger number of components ($k > 20$). In the sparse data scenarios, regardless of whether $k$ is underspecified or overspecified, adding the bias correction procedure seems to achieve either comparable or better performance. In other words, the effect of overfitting is usually alleviated by debiasing estimates from Poisson SVD.

\section{Conclusion and Discussion\label{conclusion}}
In this work, we have proposed a two-step bias correction procedure on loadings and scores estimated by Poisson SVD, and this procedure can significantly improve the accuracy of the estimates. Iterative bootstrap is performed first on loadings. It achieves an excellent performance of mitigating the estimation bias when the true latent dimension is correctly specified and still works when the number of components is misspecified. For the second step, accounting for the estimation error in loading estimates through SIMEX can further improve the score estimates when there is less concern about extreme MLEs. Otherwise, reducing MLE bias in scores through the Firth correction is shown to be more effective for sparse count data, especially when there exist rows of all zeros in the matrix.
Moreover, we believe our bias correction procedure could be extended to other exponential family data more broadly. 

There are several future directions worth investigating based on the current work. We have modeled count data using a Poisson distribution without consideration of overdispersion or underdispersion. However, overdispersion may need to be taken into account, especially when count data is extremely sparse. In addition, we have implicitly assumed that all zeros are sampling zeros. In real applications, structural zeros may also exist, which leads to zero-inflated data \citep{liu2019statistical}. For example, a user's zero listening frequency on a song may be attributed to the fact that they have not been exposed to it, as opposed to knowing the song but not interested. Then if the user was exposed to the song, the listening frequency might not be zero as in the current dataset. Exploring whether the proposed bias correction procedure is applicable to overdispersed or zero-inflated sparse count data is one direction we would like to pursue in the future. 
On the theoretical front, it is of interest to further study other statistical properties of Scholz's iterative bootstrap bias correction method in the context of low-rank modeling. 
For example, \cite{guerrier2019simulation} derived some theoretical results on the convergence rate of the variance of the classical bootstrap bias correction estimator in addition to bias.
It will be intriguing to examine if the iterative bootstrap bias correction method can produce better results than the classical bootstrap method in variance as well. 

\section*{Acknowledgements} 
We thank Lun Li for his helpful comments on the earlier version of this manuscript. This research was supported in part by the National Science Foundation Grants DMS-15-13566 and DMS-20-15490.

\bibliographystyle{apalike}
\bibliography{ref}

\appendix
\section*{Appendix}
\section{Comparison of Classical Bootstrap and Iterative Bootstrap\label{bootstrap}}
\begin{figure}[htbp]
    \centering
    \includegraphics[width = \textwidth]{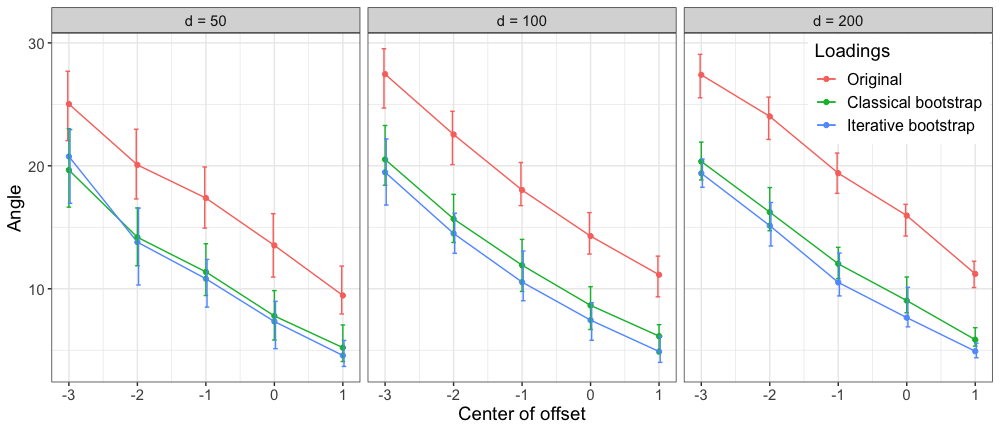}
    \caption{Comparison of angles between true loadings and estimated loadings for original estimates and bias-corrected loadings (classical bootstrap vs. iterative bootstrap). Experiments are done under $k=1$, true loadings from the standard normal distribution, different center of offsets, and $d$. The point in each bar denotes the median of replicates. The bars range from the first to third quartiles.}
    \label{fig:bootstrap}
\end{figure}

We compare two bootstrap bias correction methods for estimation of loadings. Under the classical bootstrap bias correction method, we generate 50 bootstrap samples to correct the bias. For fair comparison in terms of computational cost, we generate 10 bootstrap samples (B = 10) in the first iteration and C = 5 in the second iteration under the iterative bootstrap method. Figure~\ref{fig:bootstrap} shows the comparison of angles between estimated loadings and true loadings. Compared with the original estimated loadings from Poisson SVD, both bootstrap methods produce much smaller angles, which indicates better loading estimates. Iterative bootstrap achieves better performance than classical bootstrap especially when $d$ becomes larger.

\section{Iterative bootstrap bias correction when true rank $k$ is misspecified\label{k=2_more}}
\begin{figure}[htbp]
    \centering
    \includegraphics[width = \textwidth]{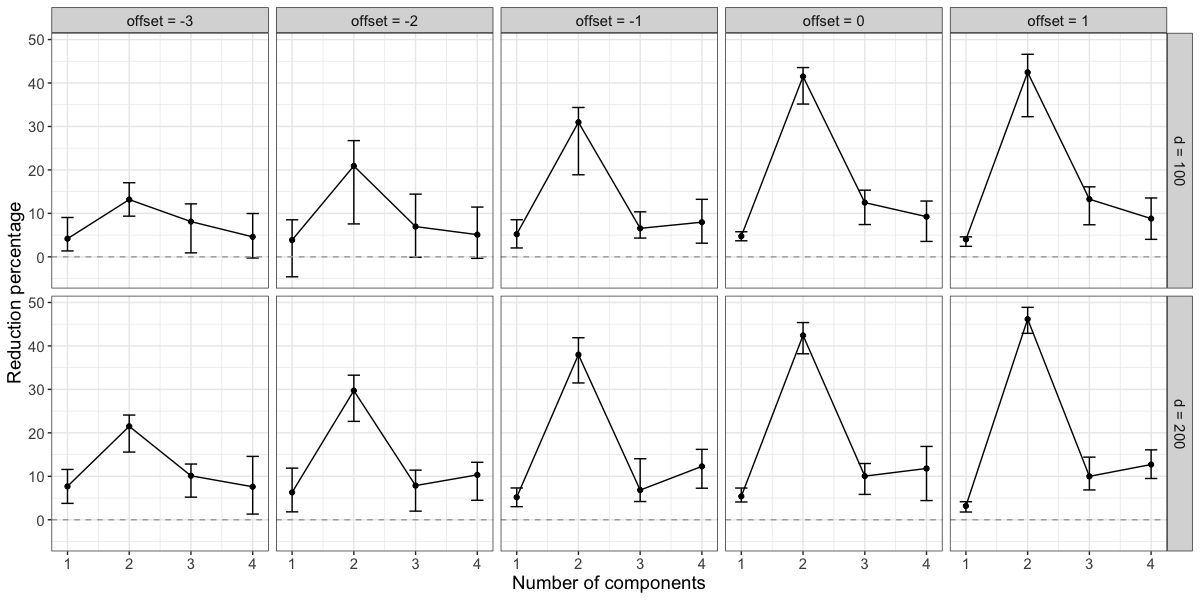}
    \caption{Percentage of reduction in RMSE of natural parameters when the true rank $k$ is misspecified under the same scenario as in Figure~\ref{k=2} except that $d = 100$ or $200$.}
    \label{fig:ib_mis_100_200}
\end{figure}

When $d = 100$ or $200$, the effect of iterative bootstrap bias correction is similar as in $d = 50$ as shown in Figure~\ref{fig:ib_mis_100_200}.
We see more benefits when $k$ is correctly specified compared to misspecified cases. 

\section{Proof of Theorem~\ref{ib_thm}}
\label{ib_proof}
The proof of Theorem~\ref{ib_thm} is provided.
\begin{proof}
By Assumption \ref{ib_assumption}, we can write the initial estimator $\hat{\bm{\pi}}$ as 
\begin{align}
\label{1}
\hat{\bm{\pi}} = \bm{\pi}_0 + \bm{c}(n)+\bm{t}(\bm{\pi}_0, n) + \epsilon(\bm{\pi}_0),
\end{align}
where $\epsilon(\bm{\pi}_0) = \hat{\bm{\pi}} - E_{Y\sim f(\pi_0)}[\hat{\bm{\pi}} ]$. 
Similarly, we can express the bootstrap estimators in the first and second iterations as
\begin{align}
\label{2}
\hat{\bm{\pi}}^{(b)} &= \hat{\bm{\pi}}+ \bm{c}(n)+\bm{t}(\hat{\bm{\pi}}, n) + \epsilon_b(\hat{\bm{\pi}}),\\
\label{3}
\hat{\bm{\pi}}^{(bc)} &= \hat{\bm{\pi}}^{(b)}+ \bm{c}(n)+\bm{t}(\hat{\bm{\pi}}^{(b)}, n) + \epsilon_{bc}(\hat{\bm{\pi}}^{(b)}), \mbox{ for } b = 1,\dots, B, c = 1,\dots, C,
\end{align}
where  $\epsilon_b(\hat{\bm{\pi}}) = \hat{\bm{\pi}}^{(b)} - E_{Y\sim f(\hat{\bm{\pi}})}[\hat{\bm{\pi}}^{(b)}]$ and $\epsilon_{bc}(\hat{\bm{\pi}}) = \hat{\bm{\pi}}^{(bc)} - E_{Y\sim f(\hat{\bm{\pi}}^{(b)})}[\hat{\bm{\pi}}^{(bc)}]$.

By using (\ref{3}), $\hat{\bm{\pi}}_{ib} = 3\hat{\bm{\pi}} - \frac{3}{B}\sum_{b=1}^B\hat{\bm{\pi}}^{(b)} + \frac{1}{B}\sum_{b=1}^B\frac{1}{C}\sum_{c=1}^C \hat{\bm{\pi}}^{(bc)}$ can be expressed first as   
\begin{align*}
    \hat{\bm{\pi}}_{ib} &= 3\hat{\bm{\pi}} - \frac{3}{B}\sum_{b=1}^B\hat{\bm{\pi}}^{(b)} + \left[\frac{1}{B}\sum_{b=1}^B\hat{\bm{\pi}}^{(b)}+ \bm{c}(n)+ \frac{1}{B}\sum_{b=1}^B\bm{t}(\hat{\bm{\pi}}^{(b)}, n) + \frac{1}{BC}\sum_{b=1}^B\sum_{c=1}^C\epsilon_{bc}(\hat{\bm{\pi}}^{(b)})\right]\\
    & = 3\hat{\bm{\pi}} - \frac{2}{B}\sum_{b=1}^B\hat{\bm{\pi}}^{(b)} + \left[\bm{c}(n)+ \frac{1}{B}\sum_{b=1}^B\bm{t}(\hat{\bm{\pi}}^{(b)}, n) + \frac{1}{BC}\sum_{b=1}^B\sum_{c=1}^C\epsilon_{bc}(\hat{\bm{\pi}}^{(b)})\right].
\end{align*}
Then, replacing $\hat{\bm{\pi}}^{(b)}$ in the second term with (\ref{2}), we have
\begin{align*}
    \hat{\bm{\pi}}_{ib} &= 3\hat{\bm{\pi}} -2\left[\hat{\bm{\pi}}+ \bm{c}(n)+\bm{t}(\hat{\bm{\pi}}, n) + \frac{1}{B}\sum_{b=1}^B\epsilon_b(\hat{\bm{\pi}})\right]\\
    & ~~~~+ \left[\bm{c}(n)+ \frac{1}{B}\sum_{b=1}^B\bm{t}(\hat{\bm{\pi}}^{(b)}, n) + \frac{1}{BC}\sum_{b=1}^B\sum_{c=1}^C\epsilon_{bc}(\hat{\bm{\pi}}^{(b)})\right]\\
    & = \hat{\bm{\pi}}-c(n)-\left[2\bm{t}(\hat{\bm{\pi}}, n)-\frac{1}{B}\sum_{b=1}^B\bm{t}(\hat{\bm{\pi}}^{(b)}, n)\right]-\left[\frac{2}{B}\sum_{b=1}^B\epsilon_b(\hat{\bm{\pi}})- \frac{1}{BC}\sum_{b=1}^B\sum_{c=1}^C\epsilon_{bc}(\hat{\bm{\pi}}^{(b)})\right].
\end{align*}
Finally, we replace $\hat{\bm{\pi}}$ in the first term with (\ref{1}) to get
\begin{align*}
\hat{\bm{\pi}}_{ib} & =[\bm{\pi}_0 + \bm{c}(n)+\bm{t}(\bm{\pi}_0, n) + \epsilon(\bm{\pi}_0)]-c(n)-\left[2\bm{t}(\hat{\bm{\pi}}, n)-\frac{1}{B}\sum_{b=1}^B\bm{t}(\hat{\bm{\pi}}^{(b)}, n)\right]\\&
\quad -\left[\frac{2}{B}\sum_{b=1}^B\epsilon_b(\hat{\bm{\pi}})- \frac{1}{BC}\sum_{b=1}^B\sum_{c=1}^C\epsilon_{bc}(\hat{\bm{\pi}}^{(b)})\right]\\
& = \bm{\pi}_0 + \left[\bm{t}(\bm{\pi}_0, n) - 2\bm{t}(\hat{\bm{\pi}}, n) + \frac{1}{B}\sum_{b=1}^B\bm{t}(\hat{\bm{\pi}}^{(b)}, n)\right] \\
&\quad + \left[\epsilon(\bm{\pi}_0) - \frac{2}{B}\sum_{b=1}^B\epsilon_b(\hat{\bm{\pi}}) + \frac{1}{BC}\sum_{b=1}^B\sum_{c=1}^C\epsilon_{bc}(\hat{\bm{\pi}}^{(b)})\right].
\end{align*}
Since all $\epsilon$ related terms have zero expectations, we have
$$
E[\hat{\bm{\pi}}_{ib}] = \bm{\pi}_0 + E\left[\bm{t}(\bm{\pi}_0, n) - \bm{t}(\hat{\bm{\pi}}, n) + \frac{1}{B}\sum_{b=1}^B\bm{t}(\hat{\bm{\pi}}^{(b)}, n) - \bm{t}(\hat{\bm{\pi}}, n)\right].
$$
Next, we consider the $j$th element of the expectation at the right side of the above equation. By Assumption \ref{ib_assumption}, 
\begin{align*}
 E \biggl\{&\bm{t}(\bm{\pi}_0, n) - \bm{t}(\hat{\bm{\pi}}, n)  + \frac{1}{B}\sum_{b=1}^B\bm{t}(\hat{\bm{\pi}}^{(b)}, n) - \bm{t}(\hat{\bm{\pi}}, n) \biggl\}_j\\
 &= E\biggl\{\sum_{i=1}^p  \frac{r_{i, j}}{n}\Bigl (\pi_{0j}-\hat{\pi}_j + \frac{1}{B}\sum_{b=1}^B \hat{\pi}_j^{(b)} - \hat{\pi}_j\Bigl )\biggl\} + \mathcal{O}(n^{-2})\\
 & = E\biggl\{\sum_{i=1}^p  \frac{r_{i, j}}{n}\Bigl (-\bm{c}_j(n)-\bm{t}_j(\bm{\pi}_0, n) - \epsilon_j(\bm{\pi}_0) + \bm{c}_j(n) + \bm{t}_j(\hat{\bm{\pi}}, n) + \frac{1}{B}\sum_{b=1}^B\epsilon_{b_j}(\hat{\bm{\pi}})\Bigl)\biggl\} + \mathcal{O}(n^{-2})\\
 & = E\biggl\{\sum_{i=1}^p  \frac{r_{i, j}}{n}\Bigl(\bm{t}_j(\hat{\bm{\pi}}, n) 
 - \bm{t}_j(\bm{\pi}_0, n)\Bigl)\biggl\} + \mathcal{O}(n^{-2})\\
 & = \sum_{i=1}^p  \frac{r_{i, j}}{n}E\biggl\{\Bigl(\bm{t}_j(\hat{\bm{\pi}}, n) 
 - \bm{t}_j(\bm{\pi}_0, n)\Bigl)\biggl\} + \mathcal{O}(n^{-2}).
\end{align*}
 Under Assumption \ref{ib_assumption}, 
\cite{guerrier2019simulation} has proved that $E\biggl\{\Bigl(\bm{t}_j(\hat{\bm{\pi}}, n) 
 - \bm{t}_j(\bm{\pi}_0, n)\Bigl)\biggl\} = \mathcal{O}(n^{-(1+\beta)})$, which implies  
 $\sum_{i=1}^p  \frac{r_{i, j}}{n}E\biggl\{\Bigl(\bm{t}_j(\hat{\bm{\pi}}, n) 
 - \bm{t}_j(\bm{\pi}_0, n)\Bigl)\biggl\}  = \mathcal{O}(n^{-(2+\beta)})$. Therefore, $E[\hat{\bm{\pi}}_{ib}] = \bm{\pi}_0 + \mathcal{O}(n^{-2})$.
\end{proof}

\end{document}